\documentclass{article}
\usepackage{spconf,amsmath,graphicx}
\usepackage{amsfonts}
\usepackage{multirow}
\usepackage{threeparttable}
\usepackage{float}
\usepackage{hyperref}
\usepackage{subfigure}
\usepackage{booktabs,setspace}
\usepackage{graphicx}
\usepackage{booktabs}
\usepackage{tabularx}
\usepackage{algorithm}
\usepackage{algorithmic}

\title{Two Stage Contextual Word Filtering for Context bias in Unified Streaming and Non-streaming Transducer}
\name{Zhanheng Yang$^{1*}$, Sining Sun$^{1*}$, Xiong Wang$^{2}$, Yike Zhang$^{2}$, Long Ma$^{2}$, Lei Xie$^{1\dagger}$}
\address{$^1$Audio, Speech and Langauge Processing Group (ASLP@NPU)\\
$^2$Tencent Technology Co., Ltd, Beijing, China \\
\small{\texttt{zhhyang@mail.nwpu.edu.cn, \{chnxwang,yikezhang,malonema\}@tencent.com,}} \\
\small{\texttt{ssning2013@gmail.com,lxie@nwpu.edu.cn}}
\thanks{$^*$Equal contribution. $^\dagger$Lei Xie is the corresponding author.}}

\begin{document}

\ninept
\maketitle
\begin{abstract}
It is difficult for an E2E ASR system to recognize words such as  entities appearing infrequently in the training data. A widely used method to mitigate this issue is feeding contextual information into the acoustic model.  Previous works have proven that a compact and accurate contextual list can boost the performance significantly. In this paper, we propose an efficient approach to obtain a high quality contextual list for a unified streaming/non-streaming based E2E model. Specifically, we make use of the phone-level streaming output to first filter the predefined contextual word list then fuse it into non-casual encoder and decoder to generate the final recognition results. Our approach improve the accuracy of the contextual ASR system and speed up the inference process. Experiments on two datasets demonstrates over 20\% CER reduction comparing to the baseline system. Meanwhile, the RTF of our system can be stabilized within 0.15 when the size of the contextual word list grows over 6,000.
	
\end{abstract}
\begin{keywords}
Speech Recognition, RNN-T, Transducer with Cascade Encoders, Context-aware Training, Attention, Contextual List Filter
\end{keywords}

\vspace{-3pt}
\section{Introduction}
\vspace{-5pt}
With the superior performance over traditional hybrid system and sufficient training data, end-to-end (E2E) automatic speech recognition (ASR) systems~\cite{graves2006connectionist,chan2016listen,graves2012sequence,dong2018speech,vaswani2017attention}, such as recurrent neural network transducer (RNN-T)~\cite{graves2012sequence,he2019streaming,zhang2021tiny} and transformer~\cite{vaswani2017attention,dong2018speech,gulati2020conformer}, have recently been widely adopted. An E2E model jointly optimizes the overall recognition pipeline during training and generates the output word sequence directly. However, one main disadvantage of such E2E ASR system is its deficiency in recognizing words which appear rarely in the paired audio-to-text training data, such as song names or person names. To solve this problem, a typical solution is to incorporate contextual information into an external LM and perform rescoring on the original acoustics posterior to generate a new biased output by adopting WFST decoding~\cite{mohri2002weighted,williams2018contextual} or some kind of fusion approach~\cite{zhao2019shallow,aleksic2015bringing,gulcehre2015using}. Such solutions are flexible for deployment and do not need to retrain the network. However, a separate rescoring module can not handle the error introduced by the upstream acoustic network adequately and often lead to an unsatisfactory bias performance.

Another feasible solution, we refer as E2E context bias, is to integrate contextual information during network training~\cite{pundak2018deep,jain2020contextual,chang2021context}, which proposes an additional bias encoder with location-aware cross attention mechanism to rescore the contextual word at training and inference with label embeddings, such as contextual LAS (C-LAS)~\cite{pundak2018deep} and context-aware transformer transducer (CATT)\cite{chang2021context}. Such architectures can integrate contextual information into acoustic model to improve the output posterior distribution corresponding to the related contextual words and fit the subsequent external LM well. While as the size of the possible contextual word list grows in real applications, the accuracy and latency of the overall system descend rapidly due to dispersive attention score and heavy attention computation. 

Moreover, in practice, we may not obtain compact and accurate contextual information in advance for some scenarios. For example, in the case of phone contact scenario, we can obtain the person names from the user address book as contextual information and form a compact contextual word list (size around one to several hundreds) for the context bias module. While in the scenario of music search, we may face a large contextual word list (size over thousands) containing popular songs or artist names. In the latter case, E2E context bias cannot work well due to the large size and low quality of the contextual word list, leading to performance degradation~\cite{pundak2018deep,chang2021context,han2022improving}.

In this paper, we address this problem by introducing a \textit{filtering} strategy to large contextual word list, aiming at alleviating recognition performance degradation as well as keeping a reasonable real time factor (RTF) at run-time.
Specifically, we propose a two-stage contextual word filter module for attention-based context bias, especially for cascaded encoder ASR framework, such as~\cite{narayanan2021cascaded}, which has shown superior performance on the recognition accuracy as well as providing a prompt feedback for users. In addition, such framework allows us to make full use of the streaming output to extract a compact and accurate contextual word list from the predefined large word list. We particularly choose \textit{phone} as the streaming modeling unit to match more possible homophonic contextual word candidates efficiently. Specifically, we deploy the two stage filter module on the phone-level streaming output to filter the predefined contextual word list. The first stage of the filter module computes the posterior sum confidence (PSC) without 
considering the phone order of occurrence to immediately eliminate most of the irrelevant candidates. The second stage further computes the sequence order confidence (SOC) on the surviving candidates to produce the final contextual word list. While come to the end, we use the final contextual word list as the input for E2E context bias module to integrate the contextual information with the output of non-streaming encoder and decoder to generate the final character-level output. With our proposed filter and context bias approach, we can gain over 20\% relative character error rate reduction (CERR) comparing to the baseline system on two datasets representing two typical application scenarios. Meanwhile, the RTF of our system can be stabilized within 0.15 when the size of the contextual word list grows over 6,000.


The rest of the paper is organized as follows. In Section 2 we first introduce our base ASR model framework modified from~\cite{narayanan2021cascaded}. Then we introduce our two stage contextual word filter module in Section 3. Section 4 presents experiment setup, results and detailed analysis Conclusions are finally drawn in Section 5.

\vspace{-7pt}
\section{Acoustic Model and Contextual Bias}
\begin{figure*}[!h]
	\vspace{-30pt}
	\centering 
	\includegraphics[width=1.0\linewidth]{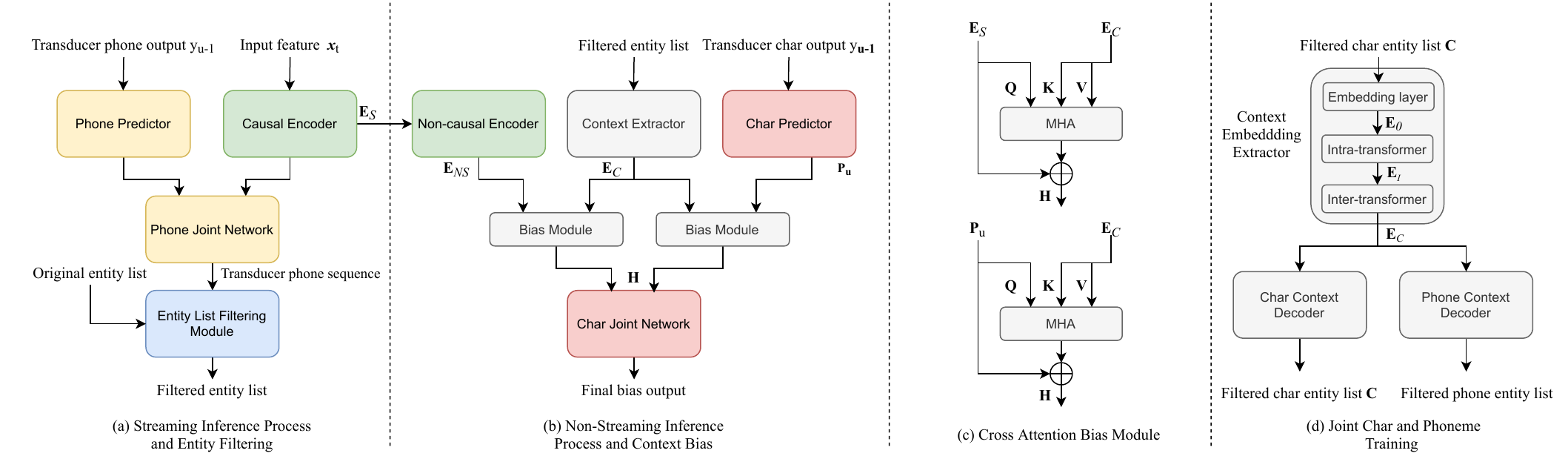}
	\caption{The proposed transducer with cascaded encoders, consisting of both streaming and non-streaming inference pipeline, i.e., [casual-encoder → phone decoder] (a) and [causal-encoder → non-causal-encoder → char decoder] (b). For the bias encoder (c), we use transducer encoder/decoder embedding as queries and $\textbf{E}_C$ as key and value for cross attention and combine the attention vector with the original input transducer encoder/decoder embedding to integrate contextual information. For joint character and phone training (d), the character and phone embeddings are learned jointly using multi-task learning.
	}\vspace{-10pt}
	\label{fig1}
\end{figure*}

\vspace{-5pt}
In this section, we introduce the overall framework of our base ASR system, as shown in Fig.~\ref{fig1}(a) and (b). Generally, we deploy an attention based E2E context bias module on the transducer model with cascaded encoders.

\subsection{System Overview}
In general, our framework is modified from~\cite{narayanan2021cascaded}. Comparing to the typical transducer, transducer with cascaded encoders consists of both causal and non-causal layers. The input features $\textbf{X}$ are first passed to a causal encoder which results in a higher-level latent representation $\textbf{E}_S$. The non-causal encoder, which is connected in cascade to the causal encoder, receives $\textbf{E}_S$ as input, and outputs $\textbf{E}_{NS}$. 

Different from~\cite{narayanan2021cascaded}, we use separate predictor and joint network for the causal and the non-causal encoders to generate modeling units of different granularity. The modeling unit of streaming and non-streaming output are phone and character respectively. There are two reasons for us to adopt different output units: 1) In our proposed approach, the streaming outputs are used to match the most possible contextual words from a big word list. Note that there are many homonyms in Chinese and phone modeling unit can match all words with the same or similar pronunciation; 2) Rare contextual words are hard to recognize for streaming model with character unit, which will affect the performance for subsequent filtering. Hence using phone as modeling unit will be better because the number of phone units is relatively small and the their coverage in training data is more comprehensive.
Same as~\cite{narayanan2021cascaded}, the total loss can be computed as the weighted sum of streaming transducer loss $L_s$ and non-streaming counterpart $L_a$:
\begin{equation}
    L_\text{Transducer} = \lambda L_s + (1 - \lambda)L_a~,
\end{equation}
where $\lambda$ is a constant weight. 

\subsection{Context Embedding Extractor}
An additional context embedding extractor is added to the non-casual encoder and the predictor as shown in Fig.~\ref{fig1}(b).  
It accepts character sequence of context words as input and generates fixed dimensional vector representations. Suppose there are $N$ contextual words. We transform words into character sequences as the input of context embedding extractor, where the maximum length of $N$ contextual words is $L$ characters. All the words are represented by characters and  padded to the same length. We use $\textbf{C}\in \{\mathbb{Z}^+\}^{N\times L}$ as the input of the extractor. Element in $\textbf{C}$ represents the integer id of a character. First, the input matrix will be transformed into an embedding matrix $\textbf{E}_0$ with shape $N\times L \times D$ using an embedding layer. Then, in order to obtain a fixed length representation for each word, a self-attention based transformer along $L$ dimension is used to transform $\textbf{E}_0$ into $\textbf{E}_1$ with shape $N\times L \times F$, where $F$ is the output dim of the transformer. We call this transformer as ``intra-transformer'' because the self-attention is calculated among a contextual word. Thanks to the self-attention,
we use the first token's transformer output as the fixed length representation for each word, where we note it as $\textbf{E}_{I}=\textbf{E}_1[:, 0, :]\in \mathbb{R}^{N\times F}$.  Furthermore, in order to model the relationship between contextual words, another transformer named ``inter-transformer'' is used along $N$ dimension of $\textbf{E}_{I}$. We obtain the final output $\textbf{E}_C$ after inter-transformer as the contextual embedding.

Moreover, we deploy a character-to-phone embedding module proposed in~\cite{chen2019joint} to assist the training of context embedding extractor. 
The module provides additional phone information to discriminate similar character sequences with different pronunciations. Thus it scales better to unseen contextual words by mapping character sequences into a space more correlated with the acoustics and better covered by the training data. Specifically, as shown in Fig.~\ref{fig1}(d), two context decoders are added to predict the corresponding character and phone sequence of the contextual words given $E_C$. The character-to-phone embedding module is learned jointly using multi-task learning. The embedding loss can be computed as:
\begin{equation}
    L_\text{Embedding} = L_\text{phone} + L_\text{char}.
\end{equation}
And the final loss of the overall system can be computed as:
\begin{equation}
    L = \alpha L_\text{Transducer} + \beta L_\text{Embedding},
\end{equation}
where $\alpha$ and $\beta$ are weight hyper-parameters.

\vspace{-5pt}
\subsection{Cross Attention Bias Module}
Given the contextual embeddings $\textbf{E}_C$, we aim to integrate contextual information into the transducer. The cross attention bias module is designed to learn relevance between contextual embeddings and utterances. The contextual embeddings will be injected into both the non-casual encoder and the predictor using attention mechanism as shown in Fig~\ref{fig1}(b) and (c). Specifically, in the cross attention bias model, the query $\textbf{Q}$ with shape $B\times T\times F$ comes from the non-casual encoder or the predictor's hidden states, where $B$, $T$ and $F$ represent batch size, sequence length and hidden dimension respectively. $\textbf{E}_C$ is used as key and value of self-attention. The final encoder or predictor hidden representations $H$ after considering contextual information can be calculated as: 

\begin{equation}
    \textbf{H} = \text{Softmax}(\frac{\textbf{QE}_C^{\top}}{\sqrt{F}})\textbf{E}_C,
\end{equation}
where $\textbf{Q}$ equals to $\textbf{E}_{NS}$ or $\textbf{P}_u$.
\section{Two Stage Contextual Word Filter}
In this section, we first introduce the overall inference pipeline of our system with our proposed two-stage contextual word filter module, as shown in Fig.~\ref{fig1}(a). Then we introduce the two-stage algorithm accordingly.

\subsection{General Inference Pipeline}
The inference process of our proposed framework can be divided into streaming and non-streaming parts. During the streaming inference, the system receives chunked acoustic feature as input and produces phone-level posterior matrix. We set a sliding window containing several output chunks for contextual word filtering. The sliding window slides forward when a new chunk output from streaming inference process comes and we use the output phone posterior matrix within the sliding window to filter the predefined contextual word list. When it comes to the non-streaming part, the filtered contextual word list is used as the context bias module input to produce the final non-streaming character sequence. We describe the two stage filtering algorithm in detail as follows.

\subsection{Posterior Sum Confidence}
The first stage compute the Posterior Sum Confidence (PSC)~\cite{liu2021rnn}. The PSC only focuses on the posterior probabilities of phones related to the contextual word and does not consider the order of occurrence. Specifically, we calculate the sum of the maximum posterior probabilities of all phones belonging to the contextual word appeared in the sliding window and normalize it with the sequence length. This stage aims to immediately eliminate most of the irrelevant contextual words.

\subsection{Sequence Order Confidence}
The second stage computes the Sequence Order Confidence (SOC). The SOC focuses on the occurrence order of the contextual word phone sequence. We traverse the whole sliding window to compute the SOC score of the remaining words and set a threshold to filter the irrelevant contextual words. In detail, we design a dynamic programming algorithm as shown in Algorithm~\ref{alg1}. Given the posterior matrix $p_{t,f}$ with $T$ frames and $F$ phone classes and reference contextual word phone sequence $U$, we compute the phone sequence order score $S_{SOC}$ among contextual words by dynamic programming using auxiliary matrix $dp$.

\begin{algorithm}[h]
    \caption{Sequence Order Confidence~(SOC)}
    \label{alg1}
    \begin{algorithmic}[1]
        \REQUIRE Input posterior matrix $p_{t,f}$, contextual word phone sequence $U$.
        \ENSURE SOC score $S_{SOC}$.
        \STATE initialize a $dp$ matrix of shape $(len(U),T)$.
        \FOR{each $i \in len(U)$}
            \FOR{each $j \in [1,T]$}
            \IF{$i = 0$}
                \IF{$j = 0$}
                    \STATE $dp[i,j] = p_{j,U_i}$
                \ELSE
                    \STATE $dp[i,j] = \text{max}(dp[i,j - 1], p_{j,U_i})$
                \ENDIF
            \ELSE
                \IF{$j = i$}
                    \STATE $dp[i,j] = dp[i - 1, j - 1] + p_{j,U_i}$
                \ELSE
                    \STATE $dp[i,j] = \text{max}(dp[i - 1, j - 1] + p_{j,U_i}, dp[i, j - 1])$
                \ENDIF
            \ENDIF
            \ENDFOR
        \ENDFOR
        \STATE $S_{SOC} = dp[-1][-1] / len(U)$
    \end{algorithmic}
\end{algorithm}

\section{Experiments}
In this section, we introduce the training corpus and test sets and describe the experimental setup for model configuration and evaluation metrics. Experimental results and analysis are also presented. 

\subsection{Corpus Details and Contextual Word List}
Our models are trained on a set of 48,000-hour anonymized Mandarin speech corpus which is collected from Tencent in-car speech assistant products. 
We test our approach on two test sets -- Contacts and Music Search. The Contacts test set contains about 1000 utterances with 970  contextual words. Most of the contextual words are person name. For Music Search test set, there are about 3000 utterances. In this case, we can not obtain exact contextual word list in advance, so a pretty big context word list is used, which contains 6253 names of the most popular songs and artists. 

\subsection{Experimental Setups}
We use 40-dim power-normalized cepstral coefficients (PNCC)~\cite{kim2016power} feature computed with a 25ms window and shifted every 10ms as input for all experiments. For the configuration of cascaded encoder based Conformer Transducer, the causal encoder consists of a convolutional downsampling layer to achieve a time reduction rate of 4 and 12 Conformer blocks. Each Conformer block consists of 4-head 256-dim multi-head attention block, CNN with kernel size 15 and 1536 linear units. The input chunk size during the streaming inference is 53 frames, which is about 480ms. Totally 20 historical chunks will be used for Conformer layer. The non-causal encoder consists of 2 Conformer blocks. The non-casual Conformer block has the similar configuration as the casual one apart from the CNN module. Non-causal convolution is used for non-casual Conformer blocks. Both the phone predictor and character predictor are a 1-layer 640-dim LSTM. The output units for phone include 211 context-independent (CI) phones and the output units for character include 6001 commonly used Chinese characters including 26 English characters.

The context embedding extractor consists of a 4-head 256-dim Transformer block with 256-dim linear units as intra-transformer and two 4-head 256-dim Transformer blocks with 512-dim linear units as inter-transformer. The bias encoder is a MHA block to integrate the contextual information. For the joint character and phone module, both the character context decoder and the phone context decoder are a Transformer block of same configuration with the intra-transformer. Such decoders receive context vector to predict the ground truth character/phone sequence to compute the joint embedding loss.

\vspace{-10pt}
\subsection{Analysis on Contextual ASR Accuracy}
\vspace{-3pt}
All experimental results on recognition accuracy are demonstrated as character error rate (CER) and relative character error rate reduction (CERR). The results are shown in Table~\ref{tab:1}. Here the baseline system is the original Transducer with cascaded encoders without a context bias module. The topline system is the same system as the baseline except an extra context bias module, which uses the ground truth contextual words as the input contextual word list for each test utterance. As shown in Table~\ref{tab:1}, the context bias module with even full contextual word list can still get a considerable CER reduction of 30\% in  Contacts scenario. While for the scenario with a large predefined contextual word list -- Music Search, the performance of the context bias module with full contextual word list declines obviously. By contrast, with our proposed two stage contextual word filtering algorithm (PSC and SOC), both scenarios can get a considerable reduction on CER, of which about 20\% on the scenario with large predefined contextual word list (Music Search) and 40\% on the scenario with small contextual word list (Contacts). The performance of the system with our proposed contextual word filter module is obviously superior to the system with full contextual word list and is impressively close to the topline system. And the performance of the two-stage filtering algorithm (PSC and SOC) is better than the one stage algorithm (PSC).

Moreover, we also evaluate the performance of the system with our proposed contextual word filter module on unmatched domain~(UD) by cross validation. Specifically, we use the contextual word list of Music Search and evaluate on Contacts set, vice versa. It shows no damage on unmatched domain result comparing to the baseline system.
\begin{table}[]
\caption{CER and CERR (\%) for different context bias strategy.}\vspace{5pt}
\label{tab:1}
\centering
\scalebox{0.83}{
\begin{tabular}{l|cc|cc}
\toprule
\multirow{2}{*}{}                                                & \multicolumn{2}{c|}{Music Search} & \multicolumn{2}{c}{Contacts} \\
                                                                 & $\text{CER} (\%)~\downarrow$          & $\text{CERR} (\%)~\uparrow$         & $\text{CER} (\%)~\downarrow$          & $\text{CERR} (\%)~\uparrow$         \\ \midrule
Baseline                                                         & 2.51          & -                 & 10.22         & -                 \\ \hline
\begin{tabular}[c]{@{}l@{}}+ Blank list\end{tabular}  & 2.65          & -5.5              & 9.99          & 2.2               \\ \hline
\begin{tabular}[c]{@{}l@{}}+ Full list\end{tabular}   & 2.52          & -0.3              & 7.13          & 30.2              \\ \hline
\begin{tabular}[c]{@{}l@{}}+ PSC\end{tabular}         & 2.07          & 17.5              & 6.19          & 39.4              \\
\begin{tabular}[c]{@{}l@{}}++ SOC\end{tabular} & 2.00          & 20.3              & 6.01          & 41.1              \\
\begin{tabular}[c]{@{}l@{}}++ SOC (UD)\end{tabular} & 2.42          & 3.6              & 9.63          & 5.7              \\ \hline
Topline                                                          & 1.66          & 33.8              & 4.81          & 52.9              \\ \bottomrule
\end{tabular}}\vspace{-10pt}
\end{table}

\vspace{-5pt}
\subsection{Analysis on Contextual Word Filtering Performance}
\vspace{-5pt}
We further study the filtering performance of our proposed algorithm. We use Entity Recall Rate (ERR) and Average List Size (ALS) to evaluate our algorithm. We define ERR as the recall rate of the ground truth entities after filtering and ALS as the average size of contextual word list after filtering. The filtering algorithm with higher ERR and smaller ALS is better, which means keeping more correct contextual words as well as suppressing the contextual word list size. As shown in Table~\ref{tab:2}, the PSC stage can filter most of the irrelevant words in the predefined contextual word list and keep a high ERR while the use of SOC further suppresses the size of the final contextual word list with negligible decline on ERR. It further proves that our proposed two-stage contextual word filtering algorithm leads to a better recognition performance.

\begin{table}[]
\caption{ERR (\%) and ALS for different contextual word filtering algorithm.}\vspace{5pt}
\label{tab:2}
\centering
\begin{tabular}{l|cc|cc}
\toprule
\multirow{2}{*}{}                                                     & \multicolumn{2}{c|}{Music Search} & \multicolumn{2}{c}{Contacts} \\
                                                                      & $\text{ERR}(\%)~\uparrow$          & $\text{ALS}~\downarrow$       & $\text{ERR} (\%)~\uparrow$          & $\text{ALS}~\downarrow$          \\ \midrule
Original List                                                         & -                 & 6253          & -                 & 972           \\ \hline
\begin{tabular}[c]{@{}l@{}}+ PSC\end{tabular}         & 96.04             & 17            & 92.88             & 12.7          \\
\begin{tabular}[c]{@{}l@{}}++ SOC\end{tabular} & 94.36             & 3.7           & 91.19             & 2.8           \\ \bottomrule
\end{tabular}\vspace{-10pt}
\end{table}

\vspace{-5pt}
\subsection{Analysis on RTF}
\vspace{-5pt}
Finally we compare the runtime RTF between the system using our proposed filter module with the one with full contextual word list as input. Specifically RTF is measured on a 2.50GHz Intel(R) Xeon(R) Platinum 8255C CPU with single thread. As shown in Table~\ref{tab:3}, comparing the RTF between the scenario of Music Search and Contacts, with the size of the input contextual word list grows, the RTF of the system with full contextual word list as input increases rapidly and becomes unacceptable. However, with the filtered contextual word list as input, the RTF of the overall system can be kept within 0.15, which is stable when the original input contextual word list grows over 6,000.

\begin{table}[]

\caption{RTF for context bias system with different contextual word filtering algorithm.}\vspace{5pt}
\label{tab:3}
\centering
\begin{tabular}{l|cc}
\toprule
\multirow{2}{*}{} & \multicolumn{2}{c}{$\text{RTF}~\downarrow$}      \\
                  & Music Search & Contacts \\ \midrule
Original List     & 4.670         & 0.196         \\ \hline
+ PSC             & 0.147        & 0.106         \\
++ SOC            & 0.149         & 0.107         \\ \hline
Test Set Duration (sec.) & 8866        & 3215         \\ \bottomrule
\end{tabular}\vspace{-10pt}
\end{table}

\vspace{-5pt}
\section{Conclusions}
\vspace{-0.5em}
In this paper, we propose a two stage contextual word filtering module for transducer with cascade encoders and set up a new contextual ASR framework. With the new framework, the system can take advantage of external contextual information such as contacts and music lists during inference. Moreover, when the size of the input contextual word list grows, with our proposed filtering module, we can alleviate the accuracy decline as well as speed up the inference process significantly. We evaluate our approach on two test sets representing different context biasing scenarios, showing over 20\% improvement on CERR comparing to the baseline system. Impressively, the RTF of our system can be stabilized within 0.15 even when the size of the input contextual word list grows over 6,000.

\bibliographystyle{IEEEbib}
\bibliography{strings,refs}

\end{document}